\documentclass[conference]{IEEEtran}
\IEEEoverridecommandlockouts
\usepackage{cite}
\usepackage{amsmath,amssymb,amsfonts}
\usepackage{algorithmic}
\usepackage{graphicx}
\usepackage{subcaption}
\usepackage{multirow}
\usepackage{textcomp}
\usepackage{xcolor}
\usepackage{soul}

\renewcommand{\baselinestretch}{0.95}

\newcommand\freefootnote[1]{%
  \let\thefootnote\relax%
  \footnotetext{#1}%
  \let\thefootnote\svthefootnote%
}

\def\BibTeX{{\rm B\kern-.05em{\sc i\kern-.025em b}\kern-.08em
    T\kern-.1667em\lower.7ex\hbox{E}\kern-.125emX}}
\begin{document}

\title{Workload-Aware Early-Stage \\ Power Delivery Network Optimization via \\Architectural Power Traces
}

\author{
    \IEEEauthorblockN{Oran Hayes\textsuperscript{†}, Maria Pantazi-Kypraiou \textsuperscript{\ddag}, Athanasios Tziouvaras\textsuperscript{\ddag}, George Stamoulis\textsuperscript{\ddag}, \\Anuj Pathania\textsuperscript{¶}, Shreejith Shanker\textsuperscript{†}, George Floros\textsuperscript{†}}
    \IEEEauthorblockA{
    \textsuperscript{†}Department of Electronic and Electrical Engineering, Trinity College Dublin, Ireland\\
    \textsuperscript{\ddag}Department of Electrical and Computer Engineering, 
University of Thessaly, Greece \\
\textsuperscript{¶}University of Amsterdam, The Netherlands\\
\{hayesor, shankers, florosg\}@tcd.ie, \{mpantazi-, attziouv, georges\}@e-ce.uth.gr, a.pathania@uva.nl
}}

\maketitle

\begin{abstract}
Power Delivery Networks (PDNs) are critical for maintaining voltage integrity in modern multiprocessor systems. Conventional early-stage PDN planning relies on static or worst-case power assumptions, often leading to over-provisioned designs and inefficient use of routing resources.
This paper proposes a workload-aware methodology for early-stage PDN optimization based on architectural power traces. Using architectural simulations, temporal power activity is captured at fine granularity and mapped to spatial power density distributions across the chip. These distributions are then translated into current demand profiles to guide PDN topology planning at tile granularity.
By incorporating realistic workload behavior, the proposed approach enables adaptive PDN resource allocation during early design stages. Experimental results demonstrate that the method achieves up to 32.94\% reduction in PDN metal area compared to conventional worst-case designs, while maintaining compliance with IR drop and electromigration constraints.
\end{abstract}


 
\section{Introduction}
Technology scaling in VLSI systems has significantly increased transistor density and switching activity, leading to higher current demand from on-chip Power Delivery Networks (PDNs). At the same time, aggressive interconnect scaling has reduced wire widths while increasing resistance, making supply voltage drop a critical design concern \cite{7838381}. 
In addition to IR drop, reliability issues such as electromigration (EM) and inductive effects, including ground bounce, impose stringent constraints on modern PDN design \cite{8482318}. To address these challenges, designers traditionally rely on techniques such as wire sizing, decoupling capacitor insertion, and topology optimization to ensure voltage integrity and reliability. However, in high-performance multiprocessor systems, PDN design has become increasingly complex due to heterogeneous architectures, workload-dependent power variations, and tight electrical and geometric constraints \cite{9377004}. \freefootnote{*This work was funded under the COIN-3D project, which has received funding from the European Union’s Horizon Europe research and innovation program under grant agreement No. 101159667.}

As a result, PDN planning has become a critical die-level challenge that must balance voltage integrity, reliability, routing resources, and thermal constraints. Early work focused on circuit-level optimization of power grid topologies to mitigate IR drop and improve reliability. Sensitivity-based techniques were proposed to selectively reinforce grid wires and reduce voltage drop while minimizing routing overhead \cite{pdnopt}, followed by analytical optimization methods addressing electrical and geometric constraints in power grid design \cite{1424171}. More recently, PDN design has been explored at higher abstraction levels, including system-level methodologies that automatically generate PDNs for multiprocessor systems using architectural and floorplan information \cite{maria}. In addition, data-driven approaches have also been proposed to predict power grid behavior and guide PDN synthesis, reducing costly iterative simulations during physical design \cite{9516776}.

In contrast, this work proposes a workload-aware methodology that leverages architectural power traces to generate spatial power density maps and guide PDN planning during early stages of the design flow. \textit{First}, architectural simulations are used to capture temporal power activity at sub-core granularity (e.g., functional units within a core) and translate it into spatial power density distributions across the chip. \textit{Second}, these power density profiles are converted into current demand estimates that enable workload-aware PDN planning at tile granularity. Finally, experimental evaluations demonstrate that the proposed approach achieves up to 32.94\% reduction in PDN metal area compared to conventional worst-case designs, while maintaining compliance with IR drop and electromigration constraints.

The remainder of this paper is organized as follows. Section~\ref{backgorund} presents the background and motivation for workload-aware PDN planning. Section~\ref{proposed} describes the proposed workload-aware methodology. Section~\ref{exp} presents the experimental setup and evaluation results. Finally, Section~\ref{cocl} concludes the paper.

\section{Background and Motivation}\label{backgorund}
\subsection{PDN Construction}
Conventional PDNs are typically implemented as regular grid structures spanning multiple metal layers, where orthogonal power stripes distribute the current from package connections to on-chip functional blocks \cite{4483978}. During the early design stages, the grid is usually sized based on worst-case current demand and allowable voltage drop constraints. Assuming a supply voltage $V_{dd}$, the current demand of a region $i$ can be approximated as
\begin{equation}
I_i = \frac{P_i}{V_{dd}},
\end{equation}
where $P_i$ denotes the power consumed in that region. For a tile of area $A_i$, the corresponding power density can be expressed as
\begin{equation}
\rho_i = \frac{P_i}{A_i}
\end{equation}
In conventional PDN planning, a worst-case value $P_{i,\mathrm{worst}}$ is often assumed uniformly or conservatively across the design, yielding
\begin{equation}
I_{i,\mathrm{worst}} = \frac{P_{i,\mathrm{worst}}}{V_{dd}}
\end{equation}
The power grid is then dimensioned such that both voltage drop and current density limits satisfy reliability constraints, i.e.,
\begin{equation}
\Delta V_i \leq \Delta V_{\max},
\end{equation}
while ensuring that wire current densities remain below electromigration (EM) limits \cite{glsvlsi}. Although this regular and guardbanded design simplifies implementation and analysis, it often leads to over-provisioned metal resources because the actual spatial power demand is typically non-uniform and workload-dependent.

\subsection{Motivational Example}
As we mentioned earlier, conventional PDN planning typically relies on worst-case current assumptions to guarantee voltage integrity under all operating conditions. As a result, PDN resources are often uniformly allocated across the chip, leading to over-provisioned power grids and inefficient use of routing resources. However, in modern multiprocessor systems, power consumption varies significantly across architectural components depending on workload behavior.

\begin{figure}[htbp]
    \centering
    \begin{subfigure}{0.47\textwidth}
        \centering
        \includegraphics[width=\linewidth]{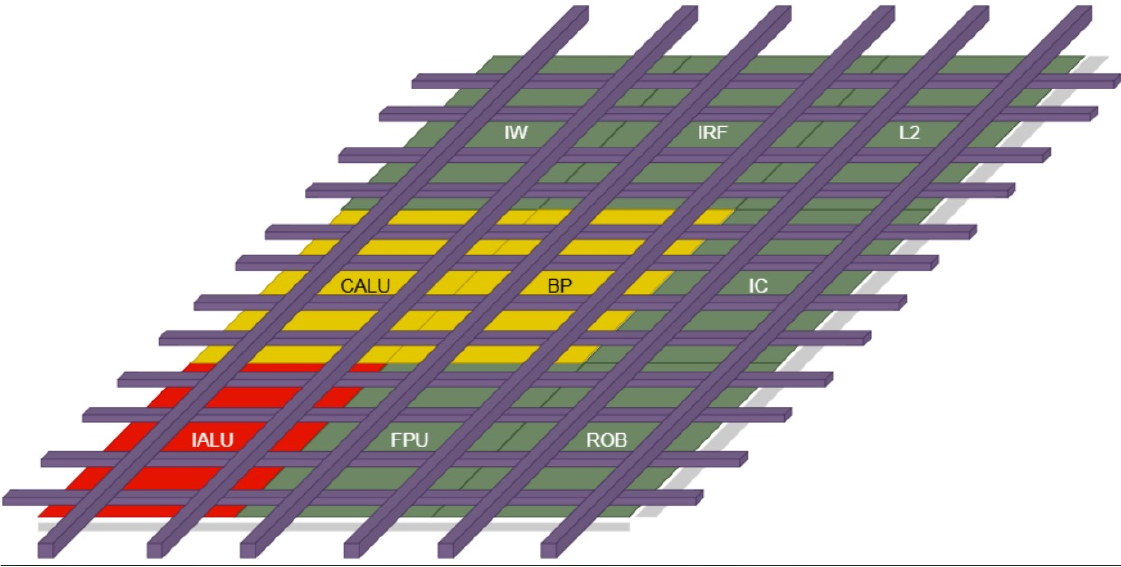}
        \caption{Uniform PDN case.}
        \label{fig:uniform}
    \end{subfigure}
    \begin{subfigure}{0.47\textwidth}
        \centering
        \includegraphics[width=\linewidth]{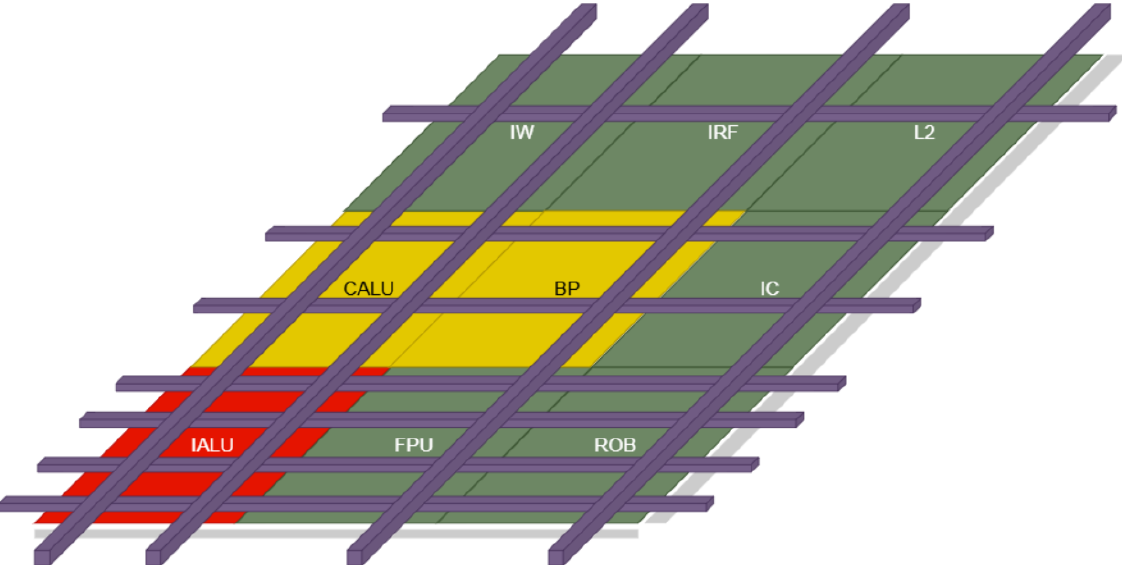}
        \caption{Optimized non-uniform PDN case.}
        \label{fig:nonuniform}
    \end{subfigure}
    
    \caption{Motivational example of PDN allocation strategies. 
(a) Conventional guardband-based PDN planning results in uniformly dense grids and over-provisioned metal resources. 
(b) Workload-aware PDN allocation adapts grid density according to spatial power demand.}
    \label{fig:comparison}
\end{figure}

More specifically, figure~\ref{fig:comparison} depicts this limitation. In conventional PDN planning (Fig.~\ref{fig:uniform}), power grid resources are uniformly distributed to satisfy worst-case current demands, resulting in a dense grid even in regions with relatively low power activity. In contrast, a workload-aware approach (Fig.~\ref{fig:nonuniform}) can adapt PDN density according to spatial power demand derived from architectural power traces, enabling more efficient allocation of metal resources while preserving voltage integrity constraints.

\section{Proposed Workload-Aware Methodology}\label{proposed}
\subsection{Architectural Power Trace Extraction}
\label{sec:powertraces}
To capture workload-dependent power behavior, architectural power traces are obtained using the HotSniper simulation
framework \cite{8444047}. HotSniper is a Sniper-based many-core simulation
toolchain that integrates architectural performance modeling
with power and thermal estimation capabilities. It enables
cycle-level simulation of multicore processors while providing
fine-grained power estimates for individual architectural
components during workload execution.

In our approach, representative workloads are executed within the HotSniper framework to collect temporal power traces at microarchitectural component-level granularity. Power consumption is monitored for major processor components, such as integer arithmetic logic units (IALUs), floating-point units (FPUs), reorder buffers (ROB), branch predictors (BP), instruction caches (IC), and other microarchitectural structures. These traces capture the dynamic variation in power consumption caused by
instruction mix, memory access behavior, and execution intensity across different architectural units. The proposed methodology generates a separate workload-aware PDN configuration for each scenario evaluated. For average-power experiments, the PDN is synthesized using temporally averaged power traces collected during execution of an individual workload. For peak power experiments, the maximum observed power per tile during workload execution is used to derive conservative current demand estimates. This enables evaluation of both workload-specific average behavior and peak-power-aware PDN provisioning.

More specifically, let $P_k(t)$ denote the instantaneous power consumption of architectural component $k$ at time step $t$. Over the
execution interval $T$, the average power for that component can be computed as:

\begin{equation}
\bar{P}_k = \frac{1}{T} \sum_{t=1}^{T} P_k(t)
\end{equation}

These component-level power traces provide a temporal representation of workload-driven power activity across the processor. Each component is mapped to its corresponding physical region in the floorplan based on architectural layout assumptions. In the next step, the extracted power information is used to derive spatial power density distributions across the chip.


\subsection{Spatial Power Density Mapping}
\label{sec:powerdensity}

The architectural power traces extracted in the previous step represent the temporal power consumption of individual
microarchitectural components. However, PDN planning requires an estimation of the spatial distribution of power demand
across the physical chip area. As a result, the architectural power information is mapped onto the chip floorplan to derive spatial power density distributions.

Let $\mathcal{K}$ denote the set of architectural components
identified in Section~\ref{sec:powertraces}. Each component $k \in \mathcal{K}$ is associated with an average power consumption $\bar{P}_k$ and occupies a physical region of area $A_k$ on the chip floorplan.

Let $\mathcal{T}$ denote the set of tiles obtained by
partitioning the chip floorplan. Each tile $i \in \mathcal{T}$
has an area $A_i$ and is associated with a subset of architectural components $\mathcal{K}_i$ that occupy that region. The total power assigned to tile $i$ is computed as

\begin{equation}
P_i = \sum_{k \in \mathcal{K}_i} \bar{P}_k 
\end{equation}

The corresponding power density is then given by

\begin{equation}
\rho_i = \frac{P_i}{A_i}
\end{equation}

Given the supply voltage $V_{dd}$, the corresponding
current demand for each tile can be estimated using

\begin{equation}
I_i = \frac{P_i}{V_{dd}}
\end{equation}

These tile-level current estimates provide a spatial representation of workload-driven current demand across
the chip. Regions exhibiting higher power density require denser PDN resources to satisfy voltage drop and EM constraints, while regions with lower power density  allow reduced grid density.

\subsection{Subcore Power Classification}

If $P_i$ denotes the power associated with tile $i$, and $P_{\max}$ represents the maximum power observed across all
tiles, then the normalized power of tile $i$ is defined as

\begin{equation}
p_i = \frac{P_i}{P_{\max}} 
\end{equation} 

Based on this normalized power metric, tiles are classified according to their relative activity level as:

\begin{equation}
C_i =
\begin{cases}
\text{High}, & T_H \le p_i \le 1 \\
\text{Medium}, & T_M \le p_i < T_H \\
\text{Low}, & 0 < p_i < T_M
\end{cases}
\end{equation}

where $T_H$ and $T_M$ are user-defined threshold parameters
that satisfy:

\begin{equation}
0 < T_M < T_H < 1
\end{equation}

These thresholds can be tuned according to the desired PDN design objectives, such as emphasizing hotspot mitigation, balancing routing overhead, or controlling overall metal allocation across the chip.









\subsection{Workload-Aware PDN Grid Construction and Current Assignment}
\label{sec:gridconstruction}

Based on the spatial power classification, the PDN grid is constructed using a workload-aware adaptive allocation
strategy. A uniform skeleton grid of candidate wire locations is first defined across the processor floorplan. Power pads are modeled as ideal voltage sources located at predefined boundary locations of the grid, while interconnect parasitics are approximated using resistive tile-level abstractions suitable for early-stage estimation.
This skeleton represents potential locations where PDN metal lines can be instantiated. The base grid spacing is
chosen such that the densest configuration is capable of supporting the maximum observed power demand $P_{\max}$.

Starting from this candidate grid, the actual PDN wires are instantiated according to the power class of each tile.
Let $k$ denote a sampling parameter that controls the spacing between inserted wires. The PDN construction procedure proceeds in three passes:

\begin{enumerate}
\item \textbf{High-power tiles:} All skeleton grid lines
intersecting high-power tiles are instantiated as PDN wires,
resulting in the maximum grid density.

\item \textbf{Medium-power tiles:} Every $k$-th skeleton
grid line intersecting medium-power tiles is instantiated.

\item \textbf{Low-power tiles:} Every $2k$-th skeleton
grid line intersecting low-power tiles is instantiated.
\end{enumerate}

This adaptive strategy increases grid density in regions
with higher current demand while reducing unnecessary routing resources in low-power areas.

After constructing the PDN topology, current sources are assigned to represent the power consumption of each tile. Using the tile power values $P_i$ obtained in the previous steps, the corresponding current demand $I_i$ is derived
from the supply voltage $V_{dd}$ as defined earlier.

Each tile is intersected by a set of PDN grid nodes. Let $N_i$ denote the number of grid intersections within tile $i$. The current demand is distributed uniformly
across these nodes such that the current injected at node $j$ is

\begin{equation}
I_{i,j} = \frac{I_i}{N_i}.
\end{equation}

This distributed current model preserves the spatial power characteristics derived from the architectural power traces while maintaining a tractable representation
for PDN analysis.

\section{Experimental Results} \label{exp}

To evaluate the proposed workload-aware PDN methodology, experiments were conducted on 4-core and 16-core processor configurations using a diverse set of multithreaded benchmarks from the PARSEC \cite{bienia2008parsec} and SPLASH-2 \cite{woo1995splash} suites. These configurations are representative of contemporary multicore processor designs, reflecting realistic system scales and workload-driven execution behavior. Simulations were performed at operating frequencies of 1GHz and 2GHz, with workloads executed at their maximum parallelism for each system to capture realistic high-concurrency behavior. The 16-core configuration focuses on SPLASH-2 benchmarks, which are well-suited for high-core-count evaluation. For each workload scenario, architectural power traces were independently collected under varying workload intensities and operating temperatures. Separate workload-aware PDN configurations were synthesized using either temporally averaged power traces or worst-case tile-level power activity. Temperature variation affects the extracted power behavior and corresponding current demand estimates used during PDN allocation. This ensures comprehensive coverage of workload-dependent power behavior, enabling robust assessment of spatial power variations and their impact on PDN resource allocation.

Moreover, in our experimental setup, the classification thresholds for workload-aware PDN allocation were defined as:
\begin{equation*}
T_H = 0.5, \qquad T_M = 0.25
\end{equation*}
These thresholds partition the chip into high-, medium-, and low-power regions based on normalized tile power, enabling adaptive allocation of PDN resources. Regions with higher activity are assigned denser power grid structures, while lower activity regions are provisioned with reduced metal resources, improving area efficiency while maintaining compliance with IR drop and electromigration constraints.

\begin{table}[h]
\centering
\caption{PDN Metal Area Reduction (\%) Using Average and Peak Power Consumption Profiles}
\label{tab:final_results}
\begin{tabular}{|c|c|c|c|}
\textbf{Cores} & \textbf{Temp (°C)} & \textbf{Avg. Power (\%)} & \textbf{Peak Power (\%)} \\
\hline
\multirow{3}{*}{4-core}
 & 0  & \textbf{32.94} & \textbf{29.05} \\
 & 40 & 32.12 & 29.05 \\
 & 65 & 32.93 & 29.05 \\
\hline
\multirow{3}{*}{16-core}
 & 0  & \textbf{32.57} & \textbf{30.31} \\
 & 40 & 32.45 & 30.31 \\
 & 65 & 30.54 & 30.31 \\

\end{tabular}
\end{table}

The results are reported in the columns of Table~\ref{tab:final_results}, where \textit{Cores} denotes the number of processing cores, \textit{Temp} represents the operating temperature (in °C), while \textit{Avg. Power} and \textit{Peak Power} indicate the percentage reduction in PDN metal area obtained using workload-aware temporally averaged tile-level power traces and maximum observed tile-level power during workload execution, respectively. All reported reductions are computed relative to a conventional uniformly provisioned PDN baseline using conservative global power assumptions.

As shown in Table~\ref{tab:final_results}, the proposed methodology achieves significant reductions in PDN metal area across all evaluated scenarios. All reported reductions are computed relative to a conventional uniform PDN grid designed using worst-case power assumptions. At the highest grid resolution, the results show up to 32.94\% reduction for the 4-core system and 32.57\% reduction for the 16-core system when using average power modeling, while the peak power model achieves up to 29.05\% and 30.31\%, respectively. Overall, the approach consistently delivers approximately 30–33\% area savings, while also highlighting that conventional worst-case modeling leads to unnecessary over-provisioning of PDN resources. These results are consistent across all evaluated workloads and operating conditions. Finally, it is important to note that all evaluated configurations satisfy IR drop and electromigration constraints without violations, confirming that the proposed reduction in PDN resources does not compromise reliability.

Furthermore, the results highlight the inherent inefficiency of conventional worst-case PDN design. By accounting for realistic workload-driven power behavior, the proposed approach avoids unnecessary over-provisioning of PDN resources, achieving up to 3.9\% additional area savings compared to the peak power allocation model. This demonstrates that workload-aware PDN planning not only improves resource efficiency but also provides a more accurate representation of spatial power demand during early design stages.

\section{Conclusions}\label{cocl}
In this paper, we presented a workload-aware methodology for early-stage PDN planning using architectural power traces. By capturing workload-dependent power behavior and mapping it to spatial power distributions, the proposed approach enables adaptive PDN resource allocation at tile granularity.
Experimental results demonstrate that the method achieves up to 32.94\% reduction in PDN metal area, while maintaining compliance with IR drop and electromigration constraints. These results highlight the limitations of conventional worst-case PDN design and demonstrate the benefits of incorporating realistic workload behavior into early-stage planning.


\bibliographystyle{IEEEtran}
\bibliography{ref}
\end{document}